# A Novel Methodology for Epidemic Risk Assessment

# of COVID-19 outbreak

A. Pluchino*[1], A.E. Biondo[2], N. Giuffrida[3], G. Inturri[4], V. Latora[1, 5, 6, 7],

R. Le Moli[8], A. Rapisarda[1, 5], G. Russo[9], C. Zappalà[1]

* Corresponding Author: alessandro.pluchino@ct.infn.it

[1] Dipartimento di Fisica e Astronomia "*Ettore Majorana*", Università di Catania and INFN Sezione di Catania, Italy;

[2] Dipartimento di Economia e Impresa, Università di Catania, Italy;

[3] Dipartimento di Ingegneria Civile e Architettura, Università di Catania, Italy;

[4] Dipartimento di Ingegneria Elettrica Elettronica e Informatica, Università di Catania, Italy;

[5] Complexity Science Hub Vienna, Austria;

[6] School of Mathematical Sciences, Queen Mary University of London, London E1 4NS, UK

[7] The Alan Turing Institute, The British Library, London NW1 2DB, UK

[8] Dipartimento di Medicina Clinica e Sperimentale - UO di Endocrinologia - Ospedale Garibaldi Nesima - Università di Catania, Italy;

[9] Dipartimento di Matematica e Informatica, Università di Catania, Italy.

**Abstract**

We propose a novel data-driven framework for assessing the *a-priori* epidemic risk of a geographical area and for identifying high-risk areas within a country. Our risk index is evaluated as a function of three different components: the hazard of the disease, the exposure of the area and the vulnerability of its inhabitants. As an application, we discuss the case of COVID-19 outbreak in Italy. We characterize each of the twenty Italian regions by using available historical data on air pollution, human mobility, winter temperature, housing concentration, health care density, population size and age. We find that the epidemic risk is higher in some of the Northern regions with respect to Central and Southern Italy. The corresponding risk index shows correlations with the available official data on the number of infected individuals, patients in intensive care and deceased patients, and can help explaining why regions such as Lombardia, Emilia-Romagna, Piemonte and Veneto have suffered much more than the rest of the country. Although the COVID-19 outbreak started in both North (Lombardia) and Central Italy (Lazio) almost at the same time, when the first cases were officially certified at the beginning of 2020, the disease has spread faster and with heavier consequences in regions with higher epidemic risk. Our framework can be extended and tested on other epidemic data, such as those on seasonal flu, and applied to other countries. We also present a policy model connected with our methodology, which might help policy-makers to take informed decisions.

# Introduction

The prediction of the future developments of a natural phenomenon is one of the main goals of science, but it remains always a great challenge when dealing with an epidemic. This proved to be particularly true in the case of the COVID-19 global pandemic that the world is suffering since January 2020.

SARS-CoV-2 is a novel coronavirus, initially announced as the causative agent of pneumonia of unknown etiology in Wuhan city, China. The genome sequence is related to a viral species named severe acute respiratory syndrome (SARS) related-CoV. These viral species also comprise some viruses detected in rhinolophid bat in Europe and Asia[1,2]. The mechanisms of immunological response to the virus infection are partially known, however a dysregulation of the immune system is very likely responsible for a worse outcome especially in patients with pre-existing respiratory or systemic diseases[3]. Most infections by coronavirus are mild and self-treated. Therefore, especially in the early stages of the disease evolution, it can be misleading to estimate the real spread of the virus just on reports of hospital and general practitioner reports. Moreover, such reports vary according to how measurements are performed, the number of tests being related very often only to the number of symptomatic patients.

Despite all this, the large amount of official data published in the last months, and updated daily[4,5], has nourished the development of several mathematical models, which are fundamental to understand the possible evolution of an epidemic and to plan effective control strategies[6-15]. However, due to the incompleteness of the data and to the intrinsic complexity of our globalized world, predicting the evolution, the peak or the end of the pandemic is a very difficult challenge [16-17]. In this paper we propose a different approach aiming, instead, at evaluating the a-priori risk of an epidemic, in particular the one caused by COVID-19. It can also result helpful in setting sound strategies to prevent or decrease the impact of future epidemic waves.

The COVID-19 outbreak started officially in China in January 2020, although probably the virus had been already circulating in the country since late October 2019 according to a recent report[18]. In Italy the first infected Italian patient was officially detected on the night of February 20 in Codogno (Lombardia), even if a recent research study of Lombardia Region reveals that more than 1000 positive cases were already present (but not tested) in that region already in the second half of January[19] or even before[20]. Moreover, at the end of January, a couple of Chinese tourists coming from Wuhan were hospitalized in Rome (Lazio) after the confirmation test of the infection. This proves that, in Italy, we have had at least two official starting points of the COVID-19 outbreak, one in the north of Italy and one in the central part[18], thus leaving some doubts about the reasons of a faster diffusion of the virus in the northern regions of Italy with respect to the central ones. Then, on March 9, a period of strict lockdown was imposed by law in order to contain the rise of the contagion. After the end of the lockdown, at the beginning of May, Italian people were able to travel again with no restrictions: most of them went to south, either to go back home or for vacation. However, this reopening of the country did not change the different and more dramatic impact that the pandemic has continued to have on different parts of Italy. In fact, at the beginning of autumn 2020, when the second epidemic wave arrived in Italy, the same northern regions (Lombardia, Emilia Romagna, Piemonte, Veneto), which have suffered more due to the first wave, seem to be still the most impacted by the pandemic with respect to the central and southern regions, in terms of severe cases and deceased persons[5].

To go more into details of this asymmetric impact of Covid-19 on the various Italian regions, let us look at the official data released by the Italian Ministry of Health[4] on April 2,  just before the epidemic peak, and on July 14, at the end of the first wave in Italy.  We report in Fig. 1a and 1b the apparent case fatality rate. It results to be quite high in northern Italian regions, about 10-14%, compared to the center and southern ones, where it is about 1-4 %. In general, these estimates are higher than those observed in other parts of the world, e.g., South Korea and Japan (1.7% and 2.6%, respectively)[21]. This can be explained by considering that the official data underestimate the correct numbers of infected people, as shown in several recent studies. For example, in March, at least 60% of infected individuals were asymptomatic, and usually difficult to detect[22]. The testing strategies adopted in Italy, especially in March-April, generally consisted in checking only people showing severe symptoms and, in particular, aged over 65. Therefore, daily official records depend very much on the number of tests done on the population, resulting in a biased sample towards aged patients.

On the other hand, also the number of officially reported deaths due to Covid-19 seems to be quite underestimated, since many aged people have most probably died in their houses or nursing houses without having been tested for Covid-19. Thus, in order to have a more reliable indicator of damages caused by SARS-CoV-2, it is convenient to look at the excess of total mortality observed in Italy with respect to the average value of past five years, as reported by official data provided by ISTAT, the Italian Institute for Statistics[23], and shown here in Fig.1c. The figure shows that the impact of the pandemic has been much more dramatic than it results from official Covid-19 data. Further, it is also clear that regions in the North of Italy have been most affected by the pandemic, even after the March-April lockdown.

The approach proposed in this paper can offer a possible explanation for the observed different diffusion and severe impact of the disease, based on a series of cofactors that differentiate the regions of Italy in various respects. In particular, the methodology we introduce, based on the Crichton's triangle[24,25], evaluates the epidemic risk index of the various Italian regions in terms of several factors, such as air pollution, people mobility, winter temperature, housing concentration, health care density, population size and age, which can be quantified using available historical data. The rationale behind the selection of these factors, as explained in the Methods section, relies on the literature, on the easy accessibility of statistical spatial data and on their uneven distribution among the Italian regions. These factors have been then combined to construct a reliable indicator of the *a-priori* epidemic risk index in Italy, which has been compared to the impact of the current COVID-19 outbreak registered in two moments: one close to the epidemic peak and the other at the end of the first epidemic wave. As we will show hereafter, Italian regions mostly affected by the pandemic (in terms of total cases, patients in intensive care units (ICU) and deceased ones) are also those with the highest risk of joining a higher propensity to spread the virus with a greater vulnerability of the population to the damage of the disease. Furthermore, we will show that our epidemic risk index fits quite well also the official available data of seasonal flu in Italy for 2019-2020[26]. Finally, we will propose a theoretical policy model, with actual examples, to design strategies aimed to the reduction of both the risk and the impact of new epidemic waves before their occurrence.

# Results

**Identification of the risk variables and their correlations with the COVID-19 damages**

We have investigated a series of factors contributing to the risk of an epidemic diffusion and its impact on the population. Among many possible, we selected the following variables: mobility index, housing concentration, healthcare density, air pollution, average winter temperature and age of population. In paragraph 1 of Methods section we motivate our choice on such variables (mainly based on epidemics literature and features of the COVID-19 outbreak), show the related data (see Table M1) and explain the adopted normalization.

The first step is, of course, to estimate to what extent the chosen normalized variables individually correlate with the main impact indicators of the COVID-19 epidemic, i.e., total cases and total deaths detected in each Italian region, cumulated up to July 14, 2020[4], when the first epidemic wave seemed to have finished, and the intensive care occupancy recorded on April 2, 2020[4], when the epidemic peak was reached. In the first two rows of Fig. 2, from panel (a) to panel (f), the spatial distributions of the six risk indicators, multiplied by the population of each region, are reported as chromatic maps and thus can be visually compared with the analogous maps of the three impact indicators, panels (g), (h) and (i) in the third row. As detailed in Table M2.1, in paragraph 2 of Methods section, pairwise correlations between risk indicators are, with a few exceptions, quite weak; furthermore, in Table M2.2, results of the linear least squares fit of each individual risk indicator to damages are reported. We found correlation coefficients ranging from 0.71 to 0.96, always higher than those observed as a function of the population, which can be considered the null model; however, the relative quadratic errors stay quite high (from 0.26 to 0.62). This suggests that some opportune combination of risk indicators could better capture the risk associated to each region. In the next paragraph, we propose a risk assessment framework aimed to this

**Definition of a risk assessment framework and calibration with COVID-19 data**

Conventional risk assessment theory relies on "Crichton's Risk Triangle"[24,25], shown in panel (l) of Fig. 2. In this framework, risk is evaluated as a function of three components: Hazard, Exposure and Vulnerability. Hazard is the potential for an event to cause harm (e.g., earthquake, flooding, epidemics); Exposure measures the amount of assets exposed to harm (e.g., buildings, infrastructures, population); Vulnerability is the harm proneness of assets if exposed to hazard events (e.g., building characteristics, drainage systems, age of population). The risk is present only when all of the three components co-exist in the same place. Used for the first time in the insurance industry[24], this approach has been extended to assess spatially distributed risks in many fields of disaster management, such as those related to climate change impact[27-31] and earthquakes[32].

In the present paper, we consider Hazard as the degree of diffusion of the virus over the population of an Italian region (influenced by a set of factors, related to spatial and socio-economic characteristics of the region itself); Exposure is the amount of people who might potentially be infected by the virus as a consequence of the Hazard (it should coincide with the size of the population of the region); Vulnerability is the propensity of an infected person to become sick or die (in general, it is strongly related to the age and pre-existing health conditions prior to infection). The combination of Vulnerability and Exposure provides a measure of the absolute damage (i.e., the number of ill people due to pathologies related to the virus in the region), which we called Consequences.

In paragraph 3 of Methods section we propose two models that differ in the way the risk indicators are aggregated into the three components of the Crichton's risk triangle. In particular, we consider the E_HV model, where the effect of Hazard and Vulnerability are combined in a single affine function of the six indicators, and the E_H_V model, where Hazard and Vulnerability are considered as affine functions of, respectively, mobility index, housing concentration and healthcare density, on one hand, and air pollution, average winter temperature and age of population on the other hand (see Fig. 2 (m) for a summary). In both models the Exposure is represented by the population of each region. Furthermore, two versions of each model have been considered: an optimized one, where the weights of the risk indicators are obtained through a least-square fitting versus real COVID-19 data, and an *a-priori* one, where all the weights are assumed to be equal.

As shown in Tables M3.1 and M3.2 of Methods section, models based on data fitting perform better, both in terms of relative mean quadratic error and correlation coefficient, as expected. In particular, the E_H_V model fits the best. Furthermore, in agreement with the strong correlation of the variables with the targets, most coefficients are positive. Indeed, all coefficients obtained by fitting the number of cases and the intensive care occupancy are positive, and only one negative coefficient appears in each model, when fitting the number of deceased. However, the numerical value of the coefficients strongly depends on both models and targets, making these models not very robust. On the other hand, the *a-priori* models are independent of the targets, depending only on the choice of the variables we decided to include in the risk evaluation.

Among the two considered *a-priori* models, where all coefficients assume the same value, we observe that the E_H_V model produces a smaller error with respect to real COVID-19 data and better correlation coefficients than the E_HV model, thus justifying the multiplicative approach which define the risk intensity in terms of the product between Hazard and Vulnerability (we used data at April 2, 2020 for this preliminary analysis but similar results would be obtained using data at July 14, 2020). Moreover, the aggregation of risk indicators in the three components of the E_H_V model follows better our motivations to choose those indicators (as explained in Methods, paragraph 1).

**Validation of the *a-priori* E_H_V Model on COVID-19 data**

Once we established the robustness of the *a-priori* E_H_V model, let us now build the corresponding regional risk ranking and validate the model with the regional COVID-19 data as a case study. In particular, following the scheme of Fig. 2 (m), by multiplying Exposure and Vulnerability for the *k*-th region, we first calculate the Consequences ($C_k = E_k \cdot V_k$, $k$=1,…,20). Then, by multiplying Hazard and Consequences, we obtain the global risk index $R_k$ for each region ($R_k = H_k \cdot C_k$, $k$=1,…, 20). In this respect, the risk index can be interpreted as the product of what is related to the occurrence of causes of the virus diffusion in a given region ($H_k$) and what is related to the severity of effects on people ($C_k$).

In Fig. 3 (a) we can appreciate the predictive capability of our model by looking at the *a-pr*iori risk ranking of the Italian regions, compared with the COVID-19 data[4], in terms of total cases (cumulated), deaths (cumulated) and intensive care occupancy (daily, not cumulated), updated both at April 2, 2020 and July 14, 2020. The values of $R_k$ have been normalized to their maximum value, so that Lombardia results to have $R_k$=1. The average of $R_k$ over all the regions is $R_{av} = 0.15$ and can be considered approximately a reference level for the Italian country (even if, of course, it has only a relative value).

As already explained, due to the intrinsic limitations of the official COVID-19 data, it is convenient to make the comparison at the aggregate level of groups of regions, without expecting to predict the exact rank within each group. Let us therefore arrange the 20 regions in four risk groups, each one characterized by a different color and ordered according to decreasing values of the risk index: very high risk ($0.4 < R_k \leq 1$, in red), high risk ($0.2 < R_k \leq 0.4$, in brown), medium risk ($0.03 < R_k \leq 0.2$, in beige) and low risk ($R_k \leq 0.03$, in pink). With this choice, our model is clearly able to correctly identify the four northern regions where the epidemic effects have been far more evident, in terms of cases, deaths and intensive care occupancy: the first in the ranking, i.e. Lombardia (whose risk score is about three times the second classified) and the group of the three regions immediately after it, Veneto, Piemonte and Emilia Romagna (even if not in the exact order of damage). A quite good agreement can be observed also for the other two groups: only for Sardegna the effects on both total cases and deaths seem to have been slightly overestimated (its insularity might play a role), while for other two regions, Umbria and Valle d'Aosta, some impact indicators have been slightly underestimated. Notice that the proposed risk classification seems quite robust, since it holds both near to the peak of April and at the end of the first wave, in July, when the intensive care occupancy of the majority of the regions was zero. In Table M3.3 reported in Methods, a further analysis of the robustness of this classification has been performed by eliminating, one by one, single indicators from the risk index definition: results show that the position of some regions slightly changes inside each group, but the composition of the four risk groups remains for the mostly unchanged with just few exceptions worsening the agreement with the impact indicators shown in Fig.3 (a). This confirms the advantage of including all indicators in the risk index.

The clear separation between northern regions from central and southern ones is also confirmed in the bottom part of Fig. 3, where the *a-priori* risk color map, in panel (c), is compared with the map of COVID-19 total cases in July, panel (b), and the map of the serious cases and deaths of the seasonal flu 2019/20 in Italy, panel (d) (ISS data[19]). The agreement is clearly visible. In Fig. 4 we show the correlations between the *a-priori* risk index and the three main impact indicators related to the outbreak, i.e. the total number of cases (a) and the total number of deaths (b), cumulated up to July 14, 2020, and the intensive care occupancy (c), registered at April 2, 2020. For each plot, a linear regression has been performed, with Pearson correlation coefficients always taking values greater or equal to 0.97, indicating a strong positive correlation. On the right of each plot we report the corresponding percentages of damage observed in the three Italian macro-regions – North, Center and South, see the geographic map (d). Also in this case the correlation is evident, if compared with the percentage of cumulated *a-priori* risk associated to the same macro-regions (e).

Another interesting way to visualize these correlations is to represent the *a-priori* risk index through its two main aggregated components, Hazard and Consequences, and plotting each region as a point of coordinates $(H_i, C_i)$ in the plane $\{H \times C\}$. This Risk Diagram is reported in Fig. 5 (a), where the points have been also characterized by the same color of the corresponding risk group of Fig. 3. It is evident that the iso-risk line described by the equation $C = R_{av} / H$ (being $R_{av} = 0.15$ the average regional risk value) is correctly able to separate the four more damaged and highly risky, northern regions (plus Lazio) from all the others. The value of the risk index is reported in parentheses next to each region name. As shown in Fig. 5 (b), where the ranking of the Italian regions has been disaggregated for both Hazard and Consequences, it is interesting to notice that some regions (such as Friuli, Trentino or Valle d'Aosta) exhibit high values of Hazard and quite low values of Consequences, while for other regions (such as Campania or Piemonte) the opposite is true. See also the colored geographic maps in Fig. 5 (c) and (d) for a visual comparison. This confirms that it is necessary to aggregate such two main components in a single global index to have a more reliable indication of the regional *a-priori* risk.

Let us close this paragraph by showing, in Fig. 6, three sequences of the geographic distribution of the total cases (a), total number of deaths (b) and current intensive care occupancy (c) as a function of time, from March 9 to July 14, 2020. These sequences are compared with the geographic map of the *a-priori* risk level (the bordered image on the right in each sequence), the latter being independent of time. In all the plots, damages seem to spread over the regions with a variable intensity (expressed by the color scale) quite correctly predicted by our *a-priori* risk analysis. The intensive care occupancy map compared with the risk map is dated April 2, since the occupancy on July 14 is zero almost everywhere (with the exception of Lombardia and a few other regions).

In the next paragraph, the methodology proposed in this paper, and in particular this representation in terms of risk diagram, will be used to build a policy model aimed at mitigating damages in case of an epidemic outbreak similar to the COVID-19 one.

**A proposal for a policy protocol to reduce the epidemic risk**

We have seen how the risk can be thought as composed in two components, one related to the causes of the infection diffusion and the other to the consequences. In this paragraph we will interpret the consequences in terms of protection and required support to people with the goal of improving the social result and/or reducing the economic cost. It is evident that enhancing the capability of the healthcare system appears to be the most important action: basically, the insufficient carrying capacity creates the emergency. Beyond specific factors explained above, the epidemic crisis in Lombardia essentially showed a breakdown of its healthcare system, caused by high demand rate for hospital admissions, long permanence times in intensive care, insufficient health assistance (diagnosis equipment, staff, spaces, etc.).

Previously illustrated data provide a *positive* analysis of an epidemic disease (i.e., how things are, in a given state of the world). The *normative* approach here described presents a viable framework to assess possible policy protocols. Several variables affecting the diffusion of an infection can be looked at as suitable policy instruments to manage both the spreading process and the stress level to the healthcare system of a given district (such as a country, a region, an urban area, etc.). Following the evidence suggested by data, we propose a theoretical model (whose details are presented in the Methods section, paragraph 4) based on two independent variables influencing the level of risk, namely the infection ratio, i.e., the proportion of infected individuals over the total population, and the number of per capita hospital beds, as a measure of the impact of consequences caused by the spreading of the disease.

We adopt an approach based on a standard model of economic policy, in which a series of instruments explicitly affecting the infection ratio and the per capita hospital beds endowment can be used to approach the target, i.e., the minimization of the risk level. A similar rationale, covering other topics, can be found in Samuelson and Solow[33] (1960) and builds upon a widely consolidated literature which dates back in time[34-39] (among many others). Despite the analysis concerns a collective problem, the model here proposed describes elements of a possible decision process followed by an individual policy-maker, thus remaining microeconomic in nature. Panel (a) in Fig. 7 shows the risk function, while the right panel provides an illustration of the family of its convex contours, for a finite set of risk levels (limited for graphic convenience):

Panel (b) in Fig. 7 replicates the meaning of Fig. 5a by translating the consequences indicated by data as the required per capita hospital beds, while explaining that the position of each iso-risk curve corresponds to the different actual composition of the scenario at hand.

We assume a unique care strategy based on the structural carrying capacity of the healthcare system, defined as the *available* number of per capita hospital beds. Such a carrying capacity derives from the health expenditure $G_H$, which is set to a level considered *sufficient*. Such a choice is based on political decisions and is reasonably inferred from past experience, structural elements of population, such as age and territorial density, etc. A part of the deliberated budget is dedicated to set up intensive care beds, as an advanced assistance service provision.

During an emergency, possibly deriving from an epidemic spreading, the number of beds can suddenly reveal insufficient. In other words, it is possible that the amount of hospital beds required at a certain point is greater than the current availability. In the model, we assume the number of hospital beds, $H$, and the proportion of intensive care beds, $\alpha$, as exogenously determined by the policy-maker who fixes $G_H$. The actual carrying capacity is shown as a function of the infection ratio, $x$, computed as the infected population over the total, as shown in panel (c) of Fig. 7, and detailed in paragraph 4 of Methods. Changes in the proportion of per capita intensive care hospital beds over the total, cause instead, a variation in the slope of the line (which becomes steeper for reduction in the proportion of intensive care beds). Finally, changes in the overall expenditure shift the line with the same slope (above for increments of the expenditure). In particular, it is worth to notice that the political choice of the ratio $\alpha = HH/H$ may imply that the overall capacity to assist the entire population is not guaranteed (i.e. the intercept on the $x$ axis might be less than 1). A direct comparison of elements contained in panels (a-b) and (c-d) of Fig. 7 provides a quick inspection of the policy problem, focused to control the epidemic spreading. The constraint should be considered as a dynamic law, but since the speed of adjustment is reasonably low, we will proceed by means of a comparative statics perspective, in which a comparison of different strategies can be presented, by starting from different, static, scenarios.

Further, by definition, an emergency challenges the usual policy settings, since the speed of damages is greater than that of policy tools. In panel (e) of Fig. 7 a hypothetic country has a given carrying capacity to sustain the risk level represented by the iso-risk curve. Without an immediate availability of funds to increase the carrying capacity, the main policy target could easily be described as the transposition of the iso-risk curve to the bottom-left: the closer the curve to the origin, the higher the satisfaction for the community. Secondly, the meaning of the relationship between the curve and the line is that until the curve touches the line, the policy maker has a sort of measure of how much the problem is out of control, given by the distance between the curve and the constraint. Third, policies may try to transpose the curve to lower levels or, equivalently, the constraint upwards (with or without modification of the slope). A minimal result is reached if both are at least tangent, as depicted in panel (f) of Fig. 7.

Whenever such a tangency condition has been reached, the highest infection rate that the given health care system can sustain has been found. Further policy actions are possible to approach a lower iso-risk curve or to save resources and/or re-allocate them differently. A policy can be considered satisfactory when any of points belonging to the arc TT' is reached, e.g. the point L. Alternative policies are neither equivalent, nor requiring the same actions, and the policy-maker has to choose actions with reference to the actual data collected by its own Country. Points F and G, although carrying the same risk level as E, still represent out-of-control positions. Different regions of the plot have a different signaling power: at point F, the infection rate is low and, thus, very difficult to be further reduced.  In such a case, for example, it would be advisable to suggest health protocols which improve people safety.  On the contrary, at point G, the infection rate is so high that a limit on social interaction easily appears to be much more urgent than medical protocols.

The right mix between a demand-side and a supply-side policy to adopt is a decision of political nature. A distinction can be made by saying that demand-side policies are devoted to reduce the number of newly infected people (by means of restrictions to movements, quarantine regulations, rules of conduct, etc.) and their effects are able to lower the iso-risk curves; supply-side policies are, instead, aimed at incrementing the carrying capacity of the system (by means of expenditure for the healthcare system, increments of dedicated personnel and intensive care beds, in-house medical protocols) and their effects can shift the constraint representing the carrying capacity of the system. Politics has, then, to decide when the risk is low *enough* or the constraint is *sufficiently* high. Specific calibration of the model will allow, in a forthcoming research, a detailed analysis of policy implications, by considering actual conditions and risk factors of specific districts, thus providing the policy-maker with a toolbox for normative directions. For instance, the model can be read to analyze differences in proposed actions in Lombardia and Veneto, and in other regions or countries.

# Discussion

We have shown how a data-driven epidemic risk analysis, accounting for a proper combination of a set of cofactors, can contribute to understand the highly inhomogeneous spread of COVID-19 in Italy during the first epidemic wave (from March 2020 to June 2020), in terms of a different a-priori risk exposure of different geographical areas. Regions such as Lombardia, Veneto, Piemonte and Emilia Romagna result indeed in the first positions of our proposed a-priori risk ranking, which consists of three main components, Hazard, Exposure and Vulnerability, related, directly or indirectly, to the probability of spreading of a virus and of its harming ability. We have evaluated these three components by using historical available data on various factors that can contribute to the territorial risk. Then, assuming the existing data are reliable, we compared our risk map with real impact indicators both close to the epidemic peak and at the end of the first epidemic wave. We are aware that the information about total number of cases can heavily be underestimated and is strictly dependent on the testing strategies. For this reason we also adopted for the comparison the total number of deaths and the intensive care occupancy. In all the cases we were able to correctly identify four groups of regions where the observed epidemic effects match with the a-priori risk level.

In the second part of the paper we then advanced a theoretical policy model that provides a decision-making toolbox to face a complex phenomenon as that of an epidemic emergency. In what follows, we provide an example illustrating the application of the model, following the steps of its practical implementation. A policy maker facing an emergency outbreak, should:

STEP 1)  Detect the current Risk Profile:
       1.a) compute the infection ratio over the population, x;
       1.b) measure the demand for hospital beds, b;
STEP 2)  Measure the current Carrying Capacity of the healthcare system (supply of hospital beds), z;
STEP 3)  Check the sustainability of the epidemic burden and assess costs of possible interventions;
STEP 4)  Apply the chosen policy;
STEP 5)  Evaluate results and, if necessary, repeat.

In order to see the procedure in action, let us now go through a numerical example with hypothetical data, which can be easily substituted with actual data of any country/region, if available.

Imagine a district with a population of 10000 people, invested by a pandemic without a known therapy. Assume, further, that the healthcare system has 2500 hospital beds (that is a very conspicuous endowment), among which 1000 are intensive care ones. Consider a first case in which 1500 persons are infected and 1200 of them present symptoms and need hospital treatments. Thus, in terms of our theoretical model presented in Methods, values are set as: $n=10000$, $H=2500$, $HH=1000$; then, $\alpha=0.4$, $h=0.6$ and $z_H=0.25$.

STEP 1) The infection ratio results, currently, to be equal to $x=1500/10000=0.15$ and the evidence suggests that, over the infected part of the population, 80% requires hospital treatments. Therefore, the actual estimate of the absorbed allowance (the demand of per capita hospital beds) is $b=1200/10000=0.12$. The point representing the country risk profile in the proposed plane hazard-consequences would, then, have coordinates $(0.15, 0.12)$, as the point A in panel g) of Fig.7.

STEP 2) The line representing the carrying capacity of the healthcare system intercepts the vertical axis to the point $z_H=0.25$ and its slope is $h=(1-\alpha)=1-1000/2500=0.6$, as the black line depicted in panel g) of Fig.7. It is worth to notice that, in this example, the healthcare system would not be able to handle infection ratios greater than 41.6%, as shown by the intercept on the horizontal axis.

STEP 3) The policy maker can see that point A can be managed by means of the current carrying capacity: it lies below the line representing the constraint. Despite this apparently encouraging result, what comes next might depend on the speed of the epidemic spreading. This model is, however, not dynamic; it uses instead a comparative static approach. Let us therefore consider two hypotheses. In case the contagion is not proceeding by involving a greater share of the population, the policy maker can reasonably decide to do nothing. Contrariwise, in case of a situation where the contagion is still in the ascending phase, the policy maker can (i) compare the pace of epidemic progression, (ii) measure the time remained before the free allowance will be used and (iii) decide, correspondingly, whether it is the case to intervene or not. In case the choice is "do nothing", STEP 4) and STEP 5) are not necessary.

Consider now a different example, in the same district as before, in which the contagion has reached 3000 persons and 2500 of them need hospital treatments. In this case, the algorithm would lead to:

STEP 1) The infection ratio results to be equal to x=3000/10000=0.3 and the actual estimate of the absorbed allowance (the demand of per capita hospital beds) is b=2500/10000=0.25. The point representing the country risk profile in the proposed plane hazard-consequences has, now, coordinates $(0.3, 0.25)$, as the point B in panel g) of Fig.7, reported also in panel h).

STEP 2) The line representing the carrying capacity of the healthcare system is the same as before, reported as the black line in panel g) and h) of Fig.7.

STEP 3) The policy maker can immediately understand that the situation is out of control, since the constraint says that the healthcare system is able to allocate no more than z=0.07 per-capita beds if the infection rate reaches x=0.3, while the situation in progress requires b=0.25 per-capita beds. This can be easily seen by comparing point B and the point on the constraint with the same abscissa of B. Since b>z, the policy maker has to decide what to do in order to satisfy the excess demand. Moreover, the situation could in principle represent a further element of diffusion of the epidemic, thus making the policy intervention more urgent. Different strategies are possible and, also in this case, the speed of contagion spreading matters in tuning opportune actions.

3.a) If, for example, the progression of the disease is decreasing or even constant, a viable decision is to allocate resources to fill the gap in terms of hospital beds. Consequently, the expenditure will be incremented by $\Delta H$, represented by a parallel shift upwards of the constraint, as the dashed line passing through point B. It is worth to notice that, in principle, the deliberated expenditure could also exceed the required gap and set the constraint to higher allowance.

3.b) If, instead, the progression of the disease is increasing, the policy maker could decide to modify also the proportion of intensive care beds ($\alpha$), in order to face the probable growth of infected persons. This would make the constraint line flatter, while new expenditure will also be required to shift it upward to reach, at least, point B, as the dotted line passing through it.

3.c) Under both 3.a and 3.b, the policy maker can comparatively consider restrictive interventions aimed to reduce the infection ratio, while deciding the expenditure to expand the endowment of hospital beds. Examples of such restrictive policies could be a forced closure of restaurants, gyms and cinemas. Such interventions would have the effect to shift the point B to the bottom-left, thus associating infection ratios with lower per capita beds requirements. Then, the cost of additive hospital beds has to be compared to social cost of restrictions, in terms of tax revenues, required subsidies, unemployment and social uncertainty. The political preference and the availability of budget funds will guide the choice. Consider that the effectiveness of such restrictive initiatives is estimated according to the presumed knowledge of the social custom in the district at hand. In the example, a society where restaurant consumptions are very frequent is very likely to respond well to a restriction of this type.

It is worth to notice that such choices, i.e., the amount of governmental expenditure in the healthcare system (influencing the vertical distance between the new and the former linear constraint), the details of the regulation imposing the restrictions, etc., are of political nature and cannot be decided by the model. They will be tailored according to political preferences of the policy maker.

STEP 4) The chosen policy is applied and the consequences are measured, while the spreading continues at its pace. Evaluation and new measurements occur and the process starts again (STEP 5).

While preserving simplicity, the model is able to depict various scenarios according to actual data and can help designing policy strategies fitting the situation at hand. In particular, elements of the model can be depicted by importing data of a district (i.e., a region, a country, etc.) and follow the presented algorithm to tailor the most adequate policy. As explained above, the political preference will guide the decision, in terms of the chosen expenditure profile (i.e., whether to change H only or also $\alpha$), and in terms of possible restrictions for society, as different lockdown strategies (e.g., more drastic but specific vs more gradual but generalized). In all cases, the forced closure of socio-economic activities will serve as an ancillary tool aimed to support a potentially insufficient endowment of hospital beds, but the actual implementation relies on the ability and preference of the Government.

In conclusion, our work is a first attempt to jointly consider different factors contributing to evaluate the a-priori epidemic risk in a geographical area. Better medical knowledge and data availability will be important to further refine and improve the proposed methodology, which could also be easily applied to other countries provided that they make the necessary information accessible. Further studies will deal also with dynamic implications, thus providing more specific intuitions according to different evolutionary paths of contagion spreading.

# Figures

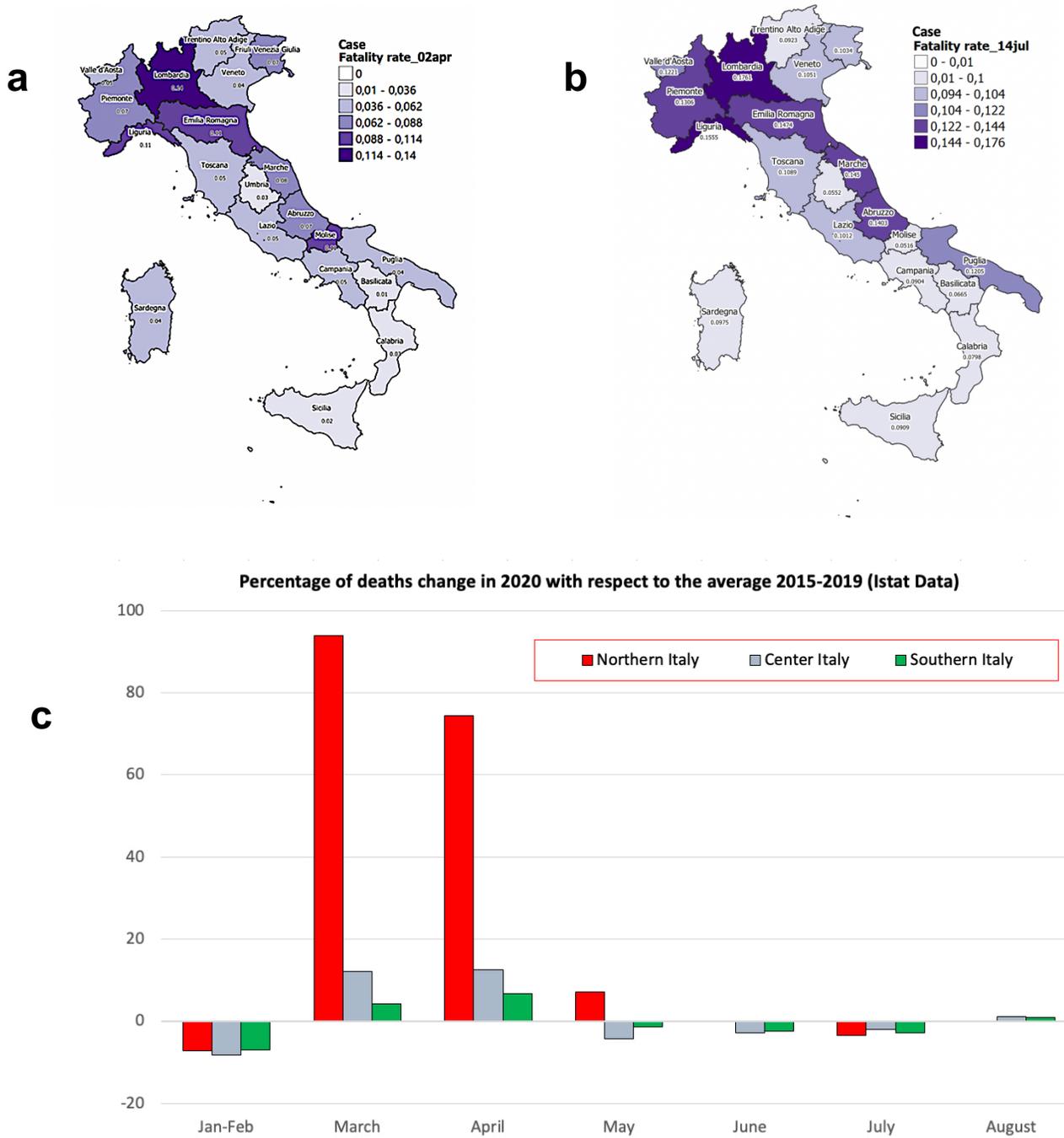

Figure 1: (a) color geographic maps of the apparent case fatality rate in the various regions on April 2, 2020 (b) and on July 14, 2020 (b), data released by the Italian Ministry of Health[4]. Maps were realized with QGIS 3.10 (https://qgis.org/en/site/). (c) Percentage of deaths change in 2020 with respect to the average taken in the period 2015-2019 (ISTAT[22]).

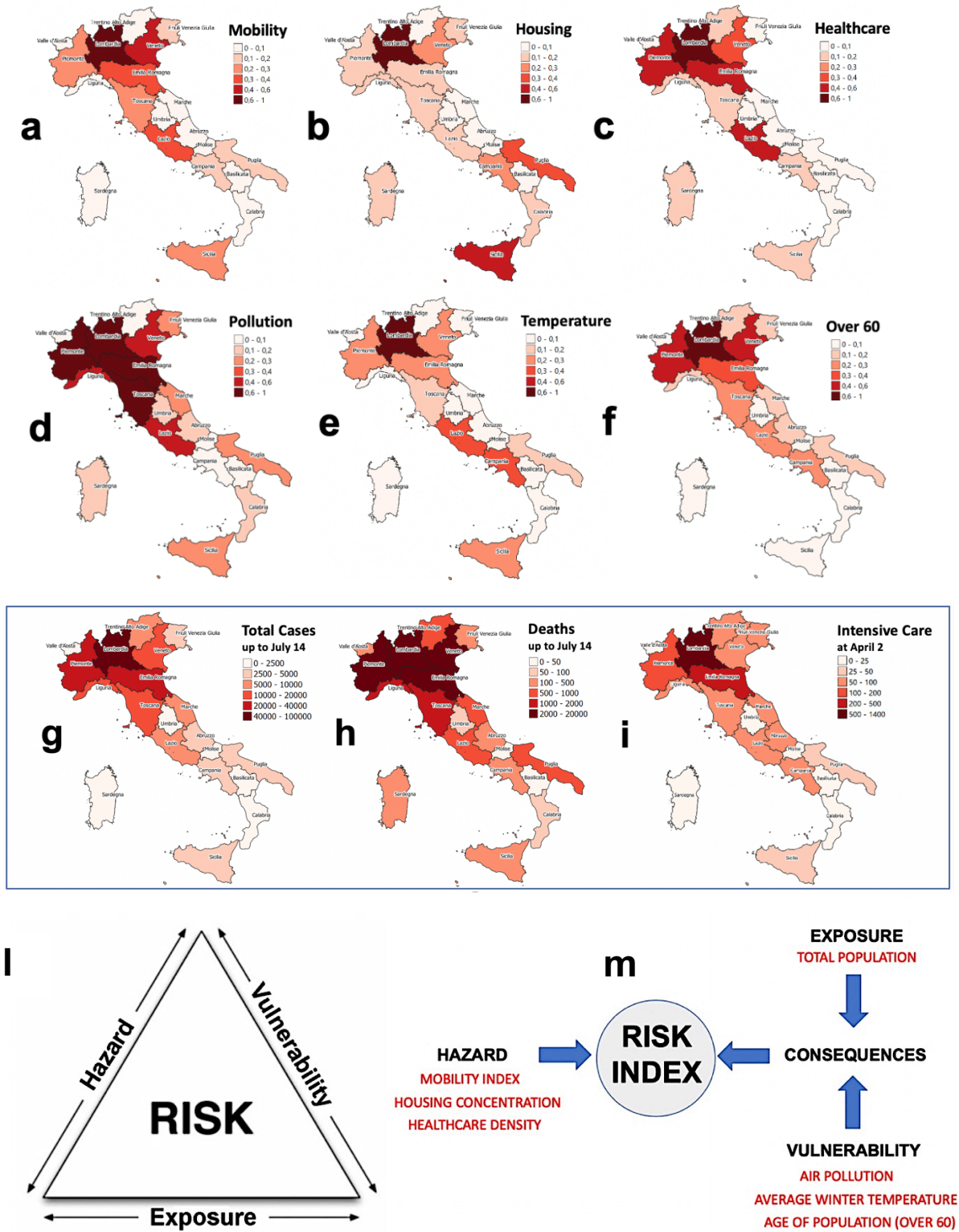

Figure 2: The geographical distribution of the six risk factors (a-f) can be compared with the COVID-19 total cases (g), the total deaths (h) and the intensive care occupancy (i). Cases and deaths have been cumulated up to July 14, 2020, i.e. at the end of the first epidemic wave; the intensive care data have been recorded on April 2, 2020, i.e. just before the epidemic peak. The risk indicators have been multiplied for the population of each region and normalized between 0 and 1 (the color scale for temperature has been reversed, i.e. dark colors mean low temperatures, see Methods). A concentration of dark colors in the northern regions is roughly visible for almost all the indicators and the correlations between the single factors and the damages range from 0.70 to 0.95. Maps were realized with QGIS 3.10 (https://qgis.org/en/site/). (l) Crichton's Risk Triangle. (m) Risk Index assessment framework: risk indicators (factors) are reported in red, risk components in black.

**a**

| # | Ranking | A-priori Risk Index | Total Cases | | Deaths | | Intensive Care | |
|---|---------|---------------------|-------------|-------------|--------|--------|----------------|-------|
| | | | 02/04/20 | 14/07/20 | 02/04/20 | 14/07/20 | 02/04/20 | 14/07/20 |
| 1 | Lombardia | 1,00 | 46065 | 95173 | 7960 | 16760 | 1351 | 27 |
| 2 | Veneto | 0,32 | 10111 | 19420 | 532 | 2041 | 345 | 2 |
| 3 | Piemonte | 0,29 | 10353 | 31507 | 983 | 4115 | 453 | 9 |
| 4 | Emilia Romagna | 0,27 | 15333 | 28971 | 1811 | 4271 | 366 | 9 |
| 5 | Lazio | 0,19 | 3433 | 8356 | 185 | 846 | 181 | 10 |
| 6 | Toscana | 0,14 | 5273 | 10330 | 268 | 1125 | 295 | 2 |
| 7 | Campania | 0,11 | 2456 | 4779 | 167 | 432 | 120 | 1 |
| 8 | Puglia | 0,09 | 2077 | 4541 | 273 | 547 | 118 | 0 |
| 9 | Friuli Venezia Giulia | 0,09 | 1799 | 3338 | 129 | 345 | 60 | 0 |
| 10 | Liguria | 0,08 | 3782 | 10038 | 488 | 1561 | 172 | 0 |
| 11 | Sicilia | 0,08 | 1791 | 3115 | 93 | 283 | 73 | 0 |
| 12 | Trentino Alto Adige | 0,05 | 3482 | 7555 | 187 | 697 | 138 | 0 |
| 13 | Marche | 0,04 | 4098 | 6805 | 503 | 987 | 164 | 0 |
| 14 | Abruzzo | 0,04 | 1497 | 3328 | 133 | 467 | 75 | 0 |
| 15 | Sardegna | 0,04 | 794 | 1374 | 40 | 134 | 24 | 0 |
| 16 | Calabria | 0,03 | 691 | 1216 | 41 | 97 | 19 | 0 |
| 17 | Umbria | 0,03 | 1128 | 1450 | 38 | 80 | 47 | 0 |
| 18 | Valle d'Aosta | 0,01 | 668 | 1196 | 63 | 146 | 25 | 0 |
| 19 | Molise | 0,01 | 165 | 446 | 11 | 23 | 8 | 0 |
| 20 | Basilicata | 0,01 | 246 | 406 | 10 | 27 | 19 | 0 |

**b** COVID-19 TOTAL CASES  **c** A-PRIORI RISK MAP  **d** SEASONAL FLU 2019-20

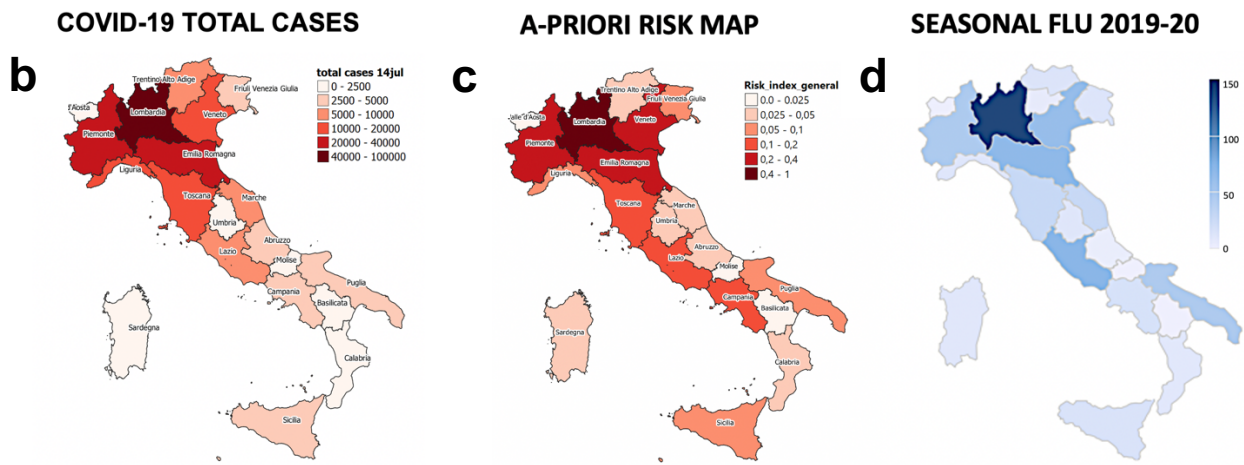

Figure 3: (a) *A-priori* normalized risk ranking of Italian regions, emerging from our analysis of risk indicators, compared with the corresponding total cases, deaths and intensive care occupancy updated, respectively, at April 2, 2020 (just before the epidemic peak) and at July 14, 2020 (at the end of the first wave). Regions are organized in four risk groups, corresponding to different colors: very high, high, medium and low risk. The agreement with the observed effects Data referring to overestimations or underestimations of risk are also colored in green and red, respectively. (b-d) Comparison between the spatial distribution of COVID-19 total cases at July 14, 2020 (b), the most struck regions (in terms of severe cases and deaths) from 2019-2020 seasonal flu (d) according to the ISS data[19] and our *a-priori* risk map (c). The geographical correlation with the risk map is evident for both kind of epidemic flus. Maps were realized with QGIS 3.10 (https://qgis.org/en/site/).

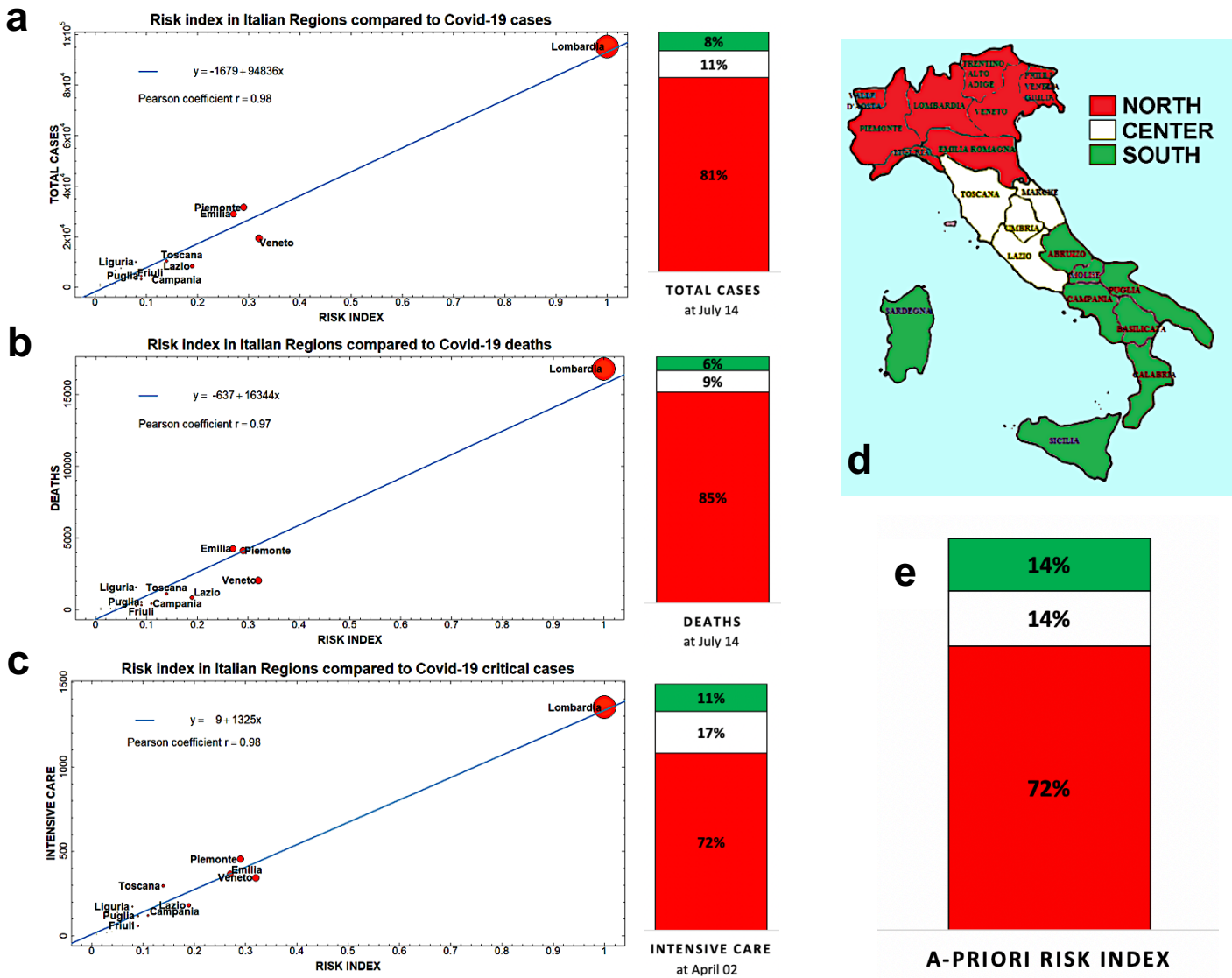

Figure 4: The three main impact indicators for COVID-19 – the total number of cases (a) and the total number of deaths (b) cumulated up to July 14, 2020[4], and the intensive care occupancy (c) at April 2, 2020[4] – are reported as function of the *a-priori* risk index for all the Italian regions. The size of the points is proportional to the risk index score. A linear regression has been performed for each plot. The Pearson correlation coefficients are very good, always greater or equal than 0.97. The corresponding percentages of damages, aggregated for the three Italian macro-regions (North, Center and South (d)) are also reported to the right and can be compared with the percentages of cumulated *a-priori* risk (e). It is clear that our *a-priori* risk index is able to explain the anomalous damage discrepancies between these different parts of Italy. Maps were realized with QGIS 3.10 (https://qgis.org/en/site/).

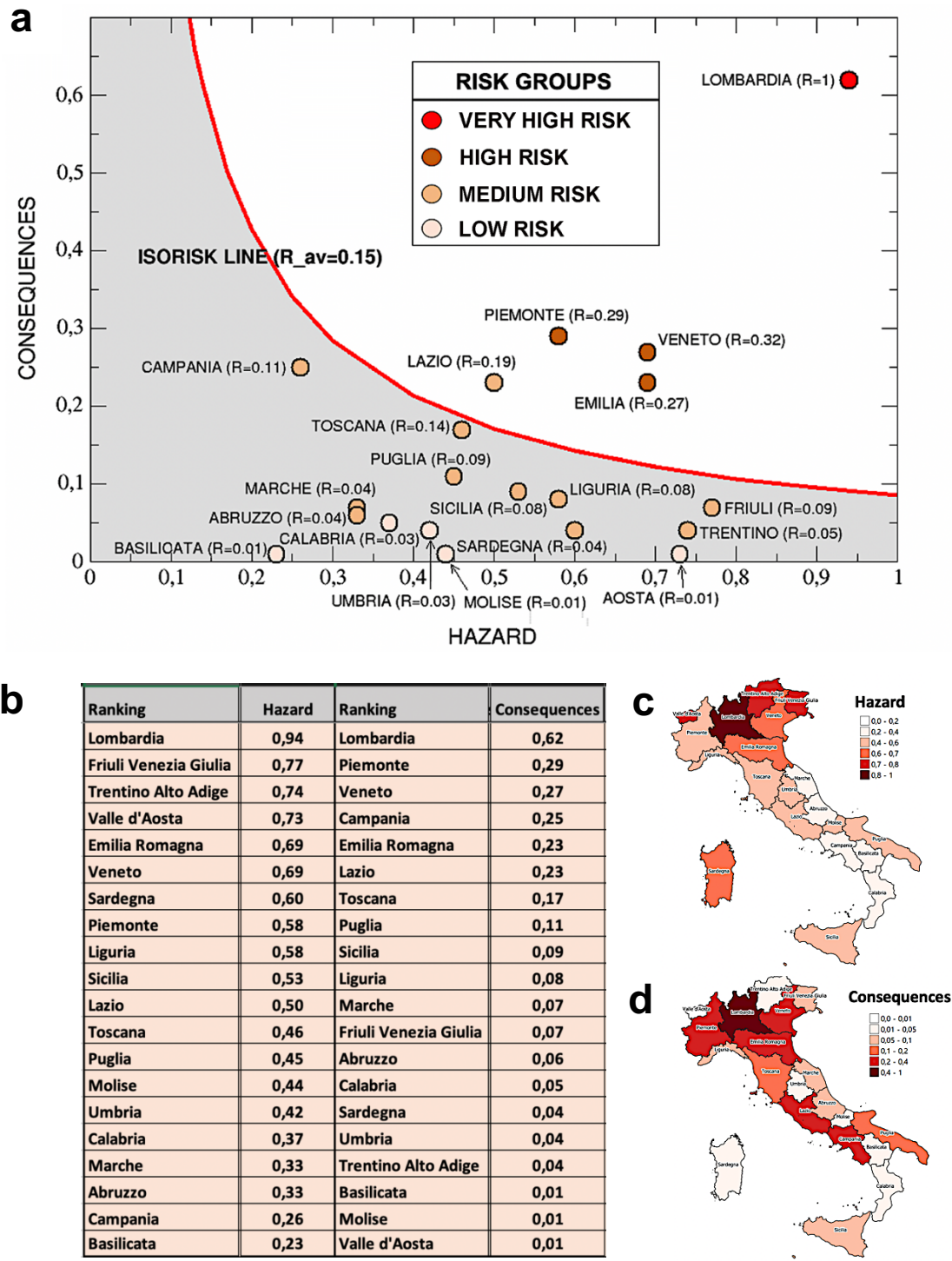

**b**

| Ranking | Hazard | Ranking | Consequences |
|---|---|---|---|
| Lombardia | 0,94 | Lombardia | 0,62 |
| Friuli Venezia Giulia | 0,77 | Piemonte | 0,29 |
| Trentino Alto Adige | 0,74 | Veneto | 0,27 |
| Valle d'Aosta | 0,73 | Campania | 0,25 |
| Emilia Romagna | 0,69 | Emilia Romagna | 0,23 |
| Veneto | 0,69 | Lazio | 0,23 |
| Sardegna | 0,60 | Toscana | 0,17 |
| Piemonte | 0,58 | Puglia | 0,11 |
| Liguria | 0,58 | Sicilia | 0,09 |
| Sicilia | 0,53 | Liguria | 0,08 |
| Lazio | 0,50 | Marche | 0,07 |
| Toscana | 0,46 | Friuli Venezia Giulia | 0,07 |
| Puglia | 0,45 | Abruzzo | 0,06 |
| Molise | 0,44 | Calabria | 0,05 |
| Umbria | 0,42 | Sardegna | 0,04 |
| Calabria | 0,37 | Umbria | 0,04 |
| Marche | 0,33 | Trentino Alto Adige | 0,04 |
| Abruzzo | 0,33 | Basilicata | 0,01 |
| Campania | 0,26 | Molise | 0,01 |
| Basilicata | 0,23 | Valle d'Aosta | 0,01 |

Figure 5: (a) Risk Diagram. Each region is represented as a point in the plane $\{H \times C\}$ while the color is proportional to the corresponding risk group updated at July 14, 2020 (see Fig.3a). The most damaged regions lie with a good approximation above the $C = R_{av}/H$ hyperbole (i.e. the iso-risk line related to the average regional risk index), while the less damaged ones lie below this line. The *a-priori* risk index score is also reported for each region. (b) The rankings of Italian regions according to either Hazard (on the left) or Consequences (on the right). The corresponding colored geographic maps are also shown in panels (c) and (d) for comparison. Maps were realized with QGIS 3.10 (https://qgis.org/en/site/).

## a. Total Cases

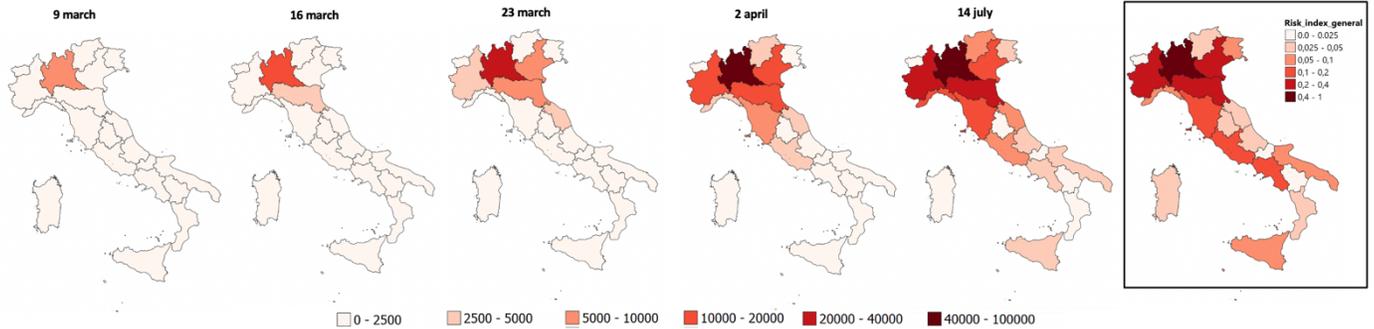

## b. Deaths

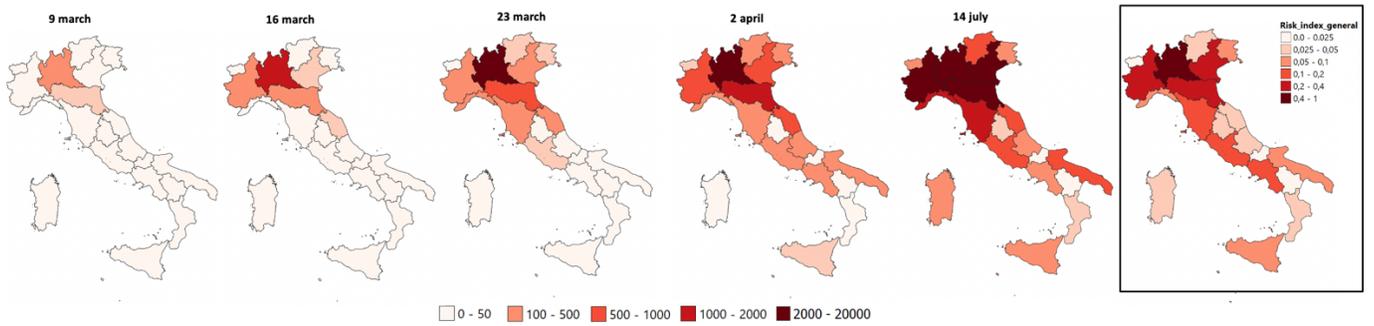

## c. Intensive Care

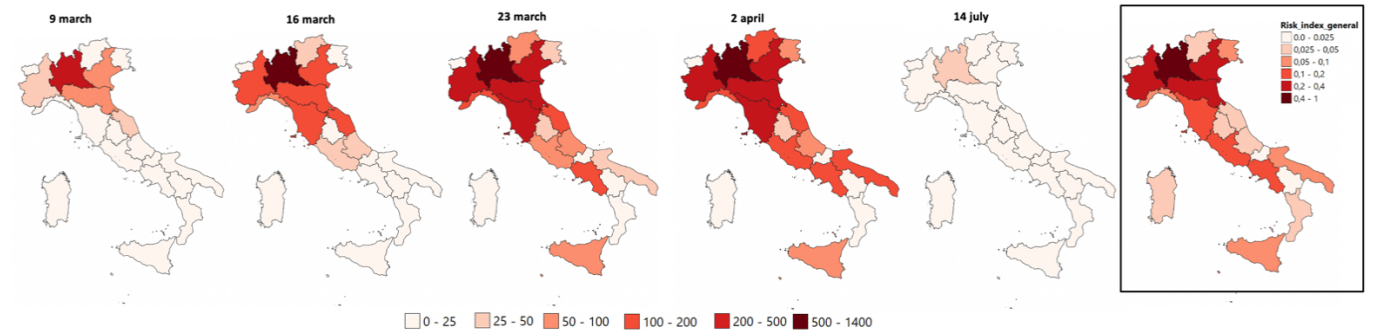

Figure 6: The geographic distributions of damage in the various Italian regions – cumulated total cases (a), cumulated total deaths (b) and daily intensive care occupancy (c) – are reported as function of time, from March 9, 2020 to July 14, 2020 and compared with the geographic distribution of the *a-priori* risk. Obviously, the intensive care occupancy to compare with the risk map is that of April, since in July, at the end the epidemic wave, this variable is zero everywhere except for a few regions (among which only Lombardia has a score slightly higher than 25). Maps were realized with QGIS 3.10 (https://qgis.org/en/site/).

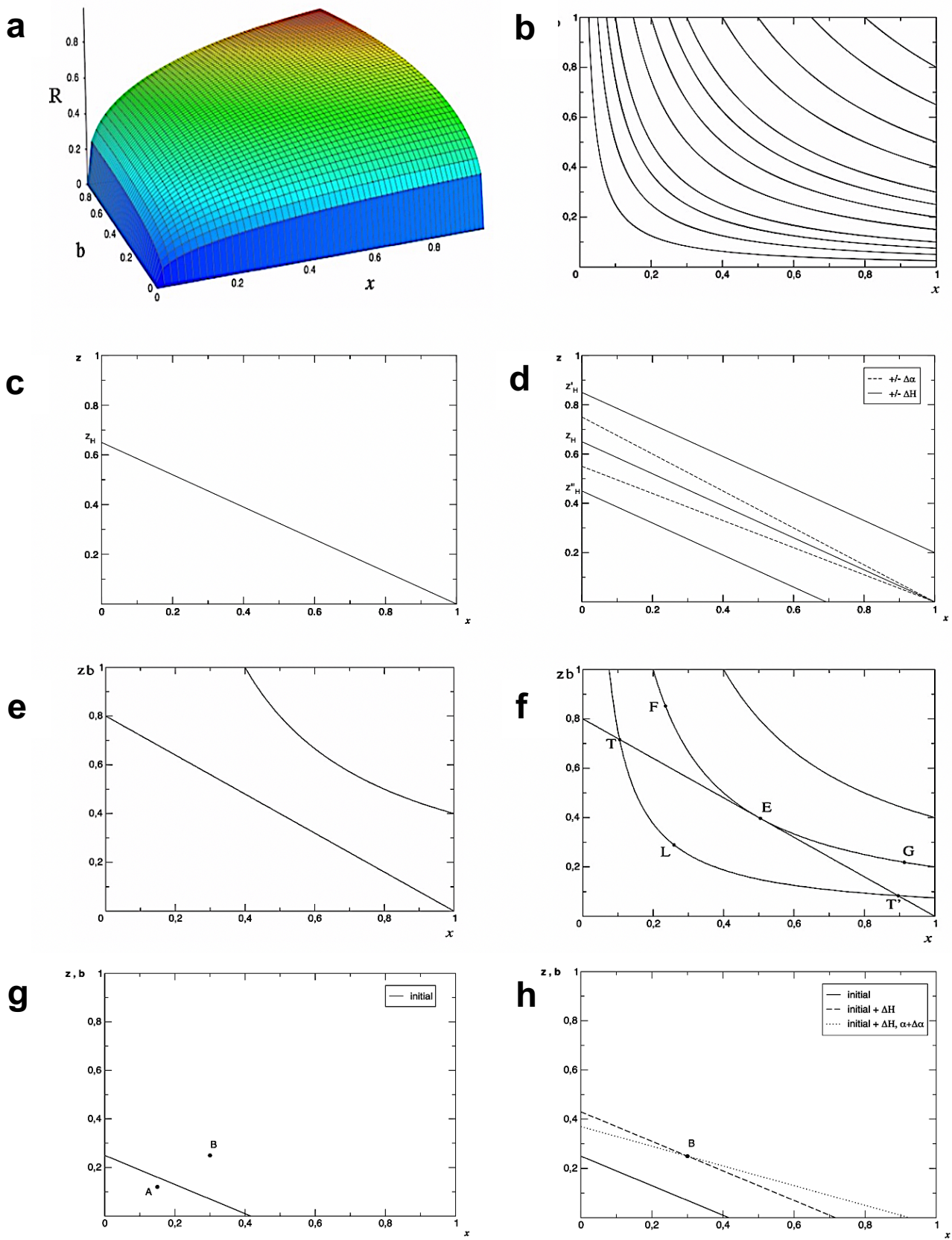

Figure 7: (a-b) The Risk function and its convex contours: an example for $R = x^{0.5}b^{0.5}$. (c-d) The carrying capacity function and effects of policy interventions on the supply-side. (e-f) Comparative statics of equilibrium and disequilibrium. (g-h) Two examples of model implementation, see the main text.

# Methods

## 1. Synthetic description of the risk indicators, population and data sources

This paragraph presents all the indicators we used, their rationale and the set of references supporting their selection. Table M1 reports the data used for each region, their definition, unit of measure and relevant source. We included in the a-priori risk index only territorial or environmental factors unevenly distributed among the regions, easily available on national databases.

**Mobility index:** Commuting data are often used to correlate population mobility and the spreading of an infection[40]. On the other hand, many recently published papers have monitored to what extent people are complying to issued travel restrictions[41] and if they are proofing to be effective in the reduction of the Covid-19 epidemic spreading[42]. According to available data, the average trip rate of mobile population in Italy is 2.50 per day and the average distance covered is 28.5 km per day[43]. We characterize each region with a "mobility index" as the regional average of the ratio between the sum of commuting flows (incoming and outgoing) for each municipality and its employed population. The data source is the Italian Ministry of Economic Policy Planning and Coordination[44]. **Housing concentration:** Urbanization increasingly affects the epidemiological characteristics of infectious disease[45,46]. Close proximity of people in their short range mobility and the attitude to use crowded public transport is amplified in compact and dense cities[47]. We capture those circumstances by the "Housing Concentration", measured as the ratio between the number of houses classified as "non detached houses" and the total number of houses. The data source is the same database cited for the mobility index. **Healthcare density:** Delayed hospital admission, misdiagnosis, unsuitable air conditioning systems, lack of infected patients segregation and inter-hospitals transfers, all might contribute in what it is commonly called super-spreading event[48]. It is worthwhile to notice that super-spreading events can occur in many situations where overcrowding in closed spaces favour the transmission of the infection. Nevertheless, our intention was to include in the a-priori risk index only territorial or environmental factors unevenly distributed among the regions, easily available on national databases. We measure the potential occurrence of this events in the infection spreading by including the "Healthcare Density" as the number of hospital beds per 10.000 inhabitants at regional level. Data source is the website of the Ministry of Health (http://www.dati.salute.gov.it/dati/dettaglioDataset.jsp?menu=dati&idPag=17). **Pollution:** Long-term exposure to air pollution may be one of the most important contributors to fatality caused by the COVID-19 in Europe[49] and in northern Italy[4]. Particulate matter (PM) is able to penetrate deeply into the respiratory tract and increase the risk of respiratory diseases[50]. According to the European Environment Agency (EEA Report No 10/2019), PM concentration in 2016 were responsible for about 374.000 premature deaths in the EU-28. Recently has been evidenced that PM10 determines a hyperactivation of JAK/STAT protein family that is associated with cells proliferation and survival[51]. JAK/STAT is also hyperactivated by several cytokines generated during the Covid-19 infection[5]. For these reasons a strengthening mechanism between PM10 and Covid-19 infection could be assumed. Besides, very recent studies are directly correlating the population exposed to particulate pollution and the contagion from COVID-19 and the consequent health damage[52,53,54]. Based on these premises, we decided to include the PM10 annual average of the mean daily concentration as a factor influencing the vulnerability of people exposed to the infection. The data source is WHO (https://www.who.int/airpollution/data/cities/en/), that provides measures at urban background, residential areas, commercial and mixed areas for the period 2013-2016. **Temperature:** Weather plays a role in the spread of 2019-nCoV[55,56,57], although not fully established. Chan[58] et al. (2011) report that a low temperature and low humidity environment may facilitate the virus transmission in subtropical areas during the spring and in air-conditioned environments. It is also commonly accepted that cold cuts down the defense barriers of the respiratory tracts[59,60]. We decided to include the average winter (from December 2016 to April 2017) daily mean temperature in each region as a factor potentially enhancing the individual vulnerability. The source of data is the Italian Ministry of Agriculture (https://www.politicheagricole.it/). **Age of population:** Most of the official data sources report more severe impacts of 2019-nCoV on elderly people, probably both for an intrinsic weakness of their immunity system and for the co-existence of other chronic pathologies. Therefore we use the ratio between the population over 60 and the total population to take into account this vulnerability factor, even if it shows only relatively small differences from one region to another. Data source is the same as population, i.e. ISTAT database (www.istat.it/it/archivio/104317). **Population:** Of course, as anyone who is infected might get ill, it is straightforward to use the total population of each region that could be affected by the infection as a measure of the risk exposure. We also use the population as a multiplying factor of each risk indicator when measuring its degree of correlation to the damage of each region (see first paragraph in the Results section). About 43% of the population is concentrated in the five regions of Northern Italy and one out of six in Lombardia. Data on regional population are available on ISTAT database (www.istat.it/it/archivio/104317).

| Indicator | Mobility inex | Housing concentration | Healthcare density | Pollution | Temperature | Age of population | Population |
|---|---|---|---|---|---|---|---|
| Data source | www.urbanindex.it | www.urbanindex.it | www.dati.salute.gov.it | www.who.int/airpollution/data/cities/en | www.politicheagricole.it | www.istat.it/it/archivio/104317 | www.istat.it/it/archivio/104317 |
| Definition | ratio between commuting flows and employed population | ratio between the number of "non detached houses" and the total number of houses | number of hospital beds per 10.000 inhabitants | annual average of PM10 daily mean concentration | average winter daily mean temperature (from December 2016 to April 2017) | ratio between over 60 population and total population | total residents living in the region |
| Region | dimensionless | dimensionless | # beds/inhab. ('0000) | $mg/m^3$ | °C | dimensionless | inha |
| Abruzzo | 0.752 | 0.871 | 33.8 | 24.3 | 5.4 | 0.281 | 1 307 309 |
| Basilicata | 0.738 | 0.869 | 32.2 | 18.7 | 8.4 | 0.267 | 578 036 |
| Calabria | 0.775 | 0.917 | 29.6 | 22.9 | 10.5 | 0.252 | 1 959 050 |
| Campania | 0.762 | 0.863 | 31.2 | 31.1 | 8.5 | 0.222 | 5 766 810 |
| Emilia Romagna | 0.823 | 0.851 | 40.1 | 24.8 | 5.7 | 0.293 | 4 342 135 |
| Friuli Venezia Giulia | 0.823 | 0.952 | 35.7 | 21.9 | 4.0 | 0.308 | 1 218 985 |
| Lazio | 0.793 | 0.834 | 37.8 | 25.3 | 7.7 | 0.265 | 5 502 886 |
| Liguria | 0.788 | 0.903 | 36.4 | 20.7 | 6.8 | 0.344 | 1 570 694 |
| Lombardia | 0.844 | 0.961 | 38.9 | 29.5 | 3.6 | 0.271 | 9 704 151 |
| Marche | 0.801 | 0.795 | 33.9 | 23.9 | 6.6 | 0.292 | 1 541 319 |
| Molise | 0.735 | 0.871 | 39.2 | 18.9 | 7.1 | 0.287 | 313 660 |
| Piemonte | 0.799 | 0.862 | 38.1 | 26.3 | 2.5 | 0.303 | 4 363 916 |
| Puglia | 0.767 | 0.950 | 31.0 | 23.2 | 9.6 | 0.253 | 4 052 566 |
| Sardegna | 0.767 | 0.962 | 35.3 | 22.4 | 10.7 | 0.266 | 1 639 362 |
| Sicilia | 0.794 | 0.942 | 31.6 | 21.7 | 11.9 | 0.250 | 5 002 904 |
| Toscana | 0.815 | 0.857 | 32.7 | 22.7 | 7.2 | 0.306 | 3 672 202 |
| Trentino Alto Adige | 0.807 | 0.888 | 40.8 | 18.1 | -1.0 | 0.247 | 1 029 475 |
| Umbria | 0.795 | 0.805 | 37.0 | 22.2 | 6.4 | 0.303 | 884 268 |
| Valle d'Aosta | 0.805 | 0.919 | 38.6 | 21.4 | -2.3 | 0.279 | 126 806 |
| Veneto | 0.838 | 0.884 | 36.1 | 27.6 | 4.3 | 0.268 | 4 857 210 |

Table M1: The risk indicators original data are reported for each Italian region, together with their sources.

## 2. Comparison between single risk indicators and impact indicators

The seven risk indicators under consideration are named below, together with their reference interval:

**Population**: $\qquad\qquad\qquad\qquad$ $X_0 \in [126806, 9704151]$

**Mobility index**: $\qquad\qquad\qquad$ $X_1 \in [0.74, 0.84]$

**Housing concentration**: $\qquad\quad$ $X_2 \in [0.80, 0.96]$

**Healthcare density**: $\qquad\qquad$ $X_3 \in [29.6, 40.80]$

**Air Pollution**: $\qquad\qquad\qquad$ $X_4 \in [18.09, 31.07]$

**Average Winter Temperature**: $\quad$ $X_5 \in [\text{-}2.29, 11.92]$

**Age of Population** (fraction of over-60 individuals): $X_6 \in [0.22, 0.34]$

These variables are suitably normalized between 0 and 1 as:

$$x_0 = \frac{X_0}{\max(X_0)} \quad ; \quad x_i = \frac{X_i - \min(X_i)}{\max(X_i) - \min(X_i)}, \quad i = 1,2,3,4,6, \quad x_5 = \frac{\max(X_5) - X_5}{\max(X_5) - \min(X_5)}$$

where $\min(X_i)$ and $\max(X_i)$ are, respectively, the minimum and the maximum value assumed by each variable $X_i$ in its own reference interval. The new normalized variables are also dimensionless.
Notice that the normalization is different for the population, since we want to avoid values equal to zero, and for the temperature, since, at variance with all other quantities, we expect that the risk increases with the decrease of the temperature.

The first test is to check possible pairwise correlations among normalized indicators, with the exception of the population (whose correlation with many other indicators is quite obvious). The Pearson correlation coefficient is reported for each couple in the correlation matrix, see Table M2.1 below.

|  | Mobility index | Housing concentration | Healthcare density | Air Pollution | Inverted temperature | Over60 concentration |
|---|---|---|---|---|---|---|
| Mobility index | 1.0000 | 0.0820 | 0.4064 | 0.3505 | 0.4640 | 0.2337 |
| Housing concentration | 0.0820 | 1.0000 | -0.1356 | -0.0510 | -0.0737 | -0.2119 |
| Healthcare density | 0.4064 | -0.1356 | 1.0000 | -0.0935 | 0.6998 | 0.3389 |
| Air Pollution | 0.3505 | -0.0510 | -0.0935 | 1.0000 | -0.0405 | -0.2527 |
| Inverted temperature | 0.4640 | -0.0737 | 0.6998 | -0.0405 | 1.0000 | 0.2183 |
| Over 60 concentration | 0.2337 | -0.2119 | 0.3389 | -0.2527 | 0.2183 | 1.0000 |

Table M2.1: Pearson correlation coefficients among the indicators $x_1, \dots, x_6$

As one can see, the majority of the indicators are weakly correlated. Noticeable exceptions concern the moderate positive correlations of some indicators, such as mobility and healthcare, with the inverted temperature. These can be explained by observing that northern Italian regions are, on average, colder and with greater mobility and healthcare density than central and southern regions.

The second thing is to check if any of the six risk indicators, $x_1, \dots, x_6$, each considered separately, can fit any of the targets $T_l, l = 1, 2, 3$, i.e. our three impact indicators: cumulative number of cases, cumulative number of deceased and number of hospitalized in intensive care at April 2, 2020. For each variable $x_i, i = 1, \dots, 6$, we consider a risk $R_{i,l} = x_0 * \alpha_{i,l} * x_i$, and each $\alpha_{i,l}$ is determined by matching the target $T_l$ in the least square sense.

In particular, we perform a linear least square fit, minimizing the following quadratic error :

$$\epsilon_{il}^2 = \sum_{k=1}^{n} \left( T_{lk} - R_{i,l} \right)^2$$

for the $i$-th risk indicator with respect to the $l$-th impact indicator. In this expression $n = 20$ is the number of regions and $T_{lk}$ denotes the impact indicator ($l = 1$, total cases, $l = 2$, number of deceased, $l = 3$, intensive care occupancy) for region $k$.

The relative mean quadratic error is defined as

$$\varepsilon_{il}^2 = \frac{\epsilon_{il}^2}{\sum_{k=1}^{n} T_{kl}^2}.$$

The result is summarized in Table M2.2. It appears that each single parameter correlates with all three targets (respectively number of cases, deceased and hospitalized in intensive care) well above the obvious correlation coefficient of the population, shown in the first column, however the correlation are not strikingly high, and the mean quadratic error is not so small.

| | | Only population | Mobility index | Housing concentr. | Healthcare density | Air pollution | Average winter temp. | Fraction over 60 |
|---|---|---|---|---|---|---|---|---|
| Relative quadratic error | Cases | 0.606 | 0.350 | 0.508 | 0.371 | 0.527 | 0.358 | 0.565 |
| | Deceased | 0.715 | 0.505 | 0.555 | 0.516 | 0.616 | 0.523 | 0.700 |
| | Intensive care | 0.533 | 0.298 | 0.481 | 0.335 | 0.480 | 0.257 | 0.471 |
| Correlation coefficient | Cases | 0.750 | 0.927 | 0.814 | 0.912 | 0.795 | 0.924 | 0.787 |
| | Deceased | 0.700 | 0.874 | 0.830 | 0.860 | 0.763 | 0.862 | 0.706 |
| | Intensive care | 0.780 | 0.934 | 0.811 | 0.914 | 0.813 | 0.954 | 0.837 |

Table M2.2: Relative mean quadratic error and correlation coefficient of single parameters.

## 3. Definition of the Risk Index and comparison among some models

The goal of this paragraph is to choose the best model of aggregation of the risk indicators presented before in the framework of the Crichton's Risk Triangle (see Results section). We observe that, in any model of this kind, the risk has to be necessarily proportional to the exposure, represented by the population. Therefore, we will assume that the risk $R$ is given by the product of Exposure $E$ times a given function $F_{HV}$ of the other parameters, related with Hazard and Vulnerability:

$$R = (E * F_{HV})$$

We propose and compare the following models, in order to understand which one is most suited for a robust risk evaluation.

1) **E_HV Linear Model.** Here the effect of Hazard and Vulnerability are combined in a single affine function of the parameters. We assume a dependence of the risk of the form:

$$R_{E\_HV} = E * F_{HV}$$

$$E = x_0 \;\; ; \;\; F_{HV} = c_{HV} + \alpha_1 x_1 + \cdots + \alpha_6 x_6$$

The coefficients $c_{HV}, \alpha_1, \ldots, \alpha_6$, in turn can be:

    a) obtained by a least square fitting;
    b) assigned *a-priori* with $c_{HV} = 0$ and all the $\alpha_i, \; i = 1, \ldots, 6$, assumed to be equal.

2) **E_H_V Multiplicative Model.** Here, Hazard and Vulnerability are considered as affine functions of, respectively, $x_1, x_2, x_3$ and $x_4, x_5, x_6$. We assume that $F_{HV}$ is the product of Hazard and Vulnerability, i.e.:

$$R_{E\_H\_V} = E * F_{HV} = E * H * V , \tag{1}$$

$$E = \alpha_0 x_0 , \tag{2}$$

$$H = c_H + \alpha_1 x_1 + \alpha_2 x_2 + \alpha_3 x_3 , \tag{3}$$

$$V = c_V + \alpha_4 x_4 + \alpha_5 x_5 + \alpha_6 x_6 . \tag{4}$$

Again, $c_H, c_V, \alpha_0, \ldots, \alpha_6$ can be:

    a) obtained by a least square fitting;
    b) assigned *a priori*, by setting $c_H = 0, c_V = 0$, and with all the $\alpha_i, \; i = 1, \ldots, 6$, assumed to be equal.

As before, we shall compare these four models (1.a, 1.b, 2.a and 2.b) versus the three types of targets available, $T_l$ ($l = 1, 2, 3$), represented by our impact indicators: cumulative number of cases, cumulative number of deceased or number of hospitalized in intensive care, all registered at April 2, 2020. In particular, for models 1.b and 2.b we adopt a linear least square fit, while, for determining the optimal parameters of models 1.a and 2.a we perform a nonlinear least square best-fit, by trying to fit the total number of cases in each region up to April 2, 2020. Since in the E_H_V model the dependence of the risk on $E, H$ and $V$ is multiplicative, we may add two normalization conditions in order to avoid infinite solutions, for example:

$$\alpha_1 + \alpha_2 + \alpha_3 = 1, \tag{5}$$

$$\alpha_4 + \alpha_5 + \alpha_6 = 1. \tag{6}$$

For all the models, we minimize the error:

$$\epsilon_l^2 = \sum_{k=1}^{n} \left( T_{lk} - E_k * F_{HV}(x_1^{(k)}, \dots, x_6^{(k)}; params) \right)^2$$

with respect to the parameters. In this expression $n = 20$ is the number of regions, $T_{lk}$ denotes the impact indicator ($l = 1$, total cases, $l = 2$, number of deceased, $l = 3$, intensive care occupancy) for region $k$, $E_k$ indicates the population of region $k$, and the function $F_{HV}$ depends on the model considered. The relative mean quadratic error $\varepsilon$ is defined as

$$\varepsilon_l^2 = \frac{\epsilon_l^2}{\sum_{k=1}^{n} T_{kl}^2}.$$

The non-linear fit is obtained by minimizing the error function by using the Levenberg-Marquardt algorithm of Matlab® (Optimization Toolbox™). Results of the best-fit with the four models are summarized in Table M3.1, while $\alpha$ coefficients of the fitting parameters, normalized so that the sum of their absolute value is 2, are reported in Table M3.2 for both E_HV (1.a) and E_H_V (2.a) models. Finally, in Table M3.3, a stability analysis of the risk ranking of the Italian regions obtained with the E_H_V *a-priori* model (1), performed by eliminating each one of the six risk indicators, $x_1, \dots, x_6$ in turn.

| | | E_HV with data fit | E_HV a-priori | E_H_V with data fit | E_H_V a-priori |
|---|---|---|---|---|---|
| Relative quadratic error | Cases | 0.15337 | 0.35737 | 0.13909 | 0.18283 |
| | Deceased | 0.19584 | 0.50521 | 0.16712 | 0.33841 |
| | Intensive care | 0.12579 | 0.28044 | 0.12025 | 0.15106 |
| Correlation coefficient | Cases | 0.98448 | 0.93505 | 0.98719 | 0.97991 |
| | Deceased | 0.97767 | 0.89278 | 0.98387 | 0.95089 |
| | Intensive care | 0.98823 | 0.94922 | 0.98923 | 0.98335 |

Table M3.1: Relative average quadratic error and correlation coefficients for E_HV and E_H_V models, for both the versions with data fitting and with *a-priori* coefficients.

|  |  | Mobility index | Housing concentr. | Healthcare density | Air pollution | Average winter temp. | Fraction over 60 |
|---|---|---|---|---|---|---|---|
|  |  | $\alpha_1$ | $\alpha_2$ | $\alpha_3$ | $\alpha_4$ | $\alpha_5$ | $\alpha_6$ |
| E_HV | Cases | 0.1965 | 0.29255 | 0.24843 | 0.064197 | 0.34052 | 0.33643 |
|  | Deceased | 0.039659 | 0.32144 | 0.2603 | -0.18589 | 0.31125 | 0.37612 |
|  | Intensive care | 0.11338 | 0.26399 | 0.06936 | 0.3598 | 0.48352 | 0.2833 |
| E_H_V | Cases | 0.32702 | 0.32792 | 0.34506 | 0.27325 | 0.47767 | 0.24908 |
|  | Deceased | -0.1015 | 0.48994 | 0.40856 | 0.16334 | 0.29603 | 0.54063 |
|  | Intensive care | 0.26213 | 0.55156 | 0.18631 | 0.38991 | 0.45037 | 0.15972 |

Table M3.2: Normalized coefficients $\alpha_i$, $i = 1, ...,6$ of both E_HV and E_V_H models, in the versions computed by fitting the data (1.a and 2.a). Most coefficients are positive, however their numerical value strongly depends on the model and on the fitted data.

| all indicators | no mobility | no housing | no healthcare | no pollution | no temp | no over 60 |
|---|---|---|---|---|---|---|
| Lombardia | Lombardia | Lombardia | Lombardia | Lombardia | Lombardia | Lombardia |
| Veneto | Piemonte | Veneto | Veneto | Piemonte | Veneto | Veneto |
| Piemonte | Veneto | Piemonte | Piemonte | Emilia Romagna | Piemonte | Piemonte |
| Emilia Romagna | Emilia Romagna | Emilia Romagna | Emilia Romagna | Veneto | Emilia Romagna | Emilia Romagna |
| Lazio | Lazio | Lazio | Toscana | Lazio | Lazio | Lazio |
| Toscana | Campania | Toscana | Lazio | Toscana | Toscana | Campania |
| Campania | Puglia | Campania | Campania | Liguria | Campania | Toscana |
| Puglia | Toscana | Liguria | Puglia | Friuli Venezia Giulia | Sicilia | Puglia |
| Friuli Venezia Giulia | Liguria | Friuli Venezia Giulia | Friuli Venezia Giulia | Trentino Alto Adige | Puglia | Friuli Venezia Giulia |
| Liguria | Friuli Venezia Giulia | Marche | Friuli Venezia Giulia | Puglia | Liguria | Sicilia |
| Sicilia | Sicilia | Trentino Alto Adige | Sicilia | Sicilia | Friuli Venezia Giulia | Trentino Alto Adige |
| Trentino Alto Adige | Sardegna | Sicilia | Sardegna | Marche | Sardegna | Liguria |
| Marche | Trentino Alto Adige | Umbria | Calabria | Sardegna | Marche | Marche |
| Abruzzo | Abruzzo | Puglia | Trentino Alto Adige | Abruzzo | Calabria | Sardegna |
| Sardegna | Calabria | Abruzzo | Marche | Umbria | Abruzzo | Abruzzo |
| Calabria | Umbria | Calabria | Abruzzo | Campania | Umbria | Calabria |
| Umbria | Marche | Sardegna | Umbria | Calabria | Trentino Alto Adige | Umbria |
| Valle d'Aosta | Molise | Valle d'Aosta | Valle d'Aosta | Valle d'Aosta | Molise | Valle d'Aosta |
| Molise | Valle d'Aosta | Molise | Basilicata | Molise | Valle d'Aosta | Molise |
| Basilicata | Basilicata | Basilicata | Molise | Basilicata | Basilicata | Basilicata |

Table M3.3: Robustness of the *a-priori* risk ranking, shown in Fig.3a of the main text, under elimination of single indicators from the risk index definition. The composition of the four risk groups seems to remain mostly unchanged. The few regions that change group have been colored in red. Notice that all these small changes worsen the group composition, in terms of the comparison with the impact indicators shown in Fig.3(a). Only the elimination of the over-60 indicator leaves the groups composition unchanged, probably due to the fact that the fraction of over 60 individuals shows only small fluctuations going from one region to another (see Table M1).

## 4. The theoretical model for policy assessment

We propose a theoretical framework to discuss the policy problem. The risk of a community has been described above as depending on several components. We now adopt a simplification and consider the whole set of possible elements reconciled on two aspects, namely the proportion of infected individuals over the total population, which we call here infection ratio, $x$, and the impact of consequences caused by the spreading of the disease, measured as the number of per capita hospital beds, $b$, required by the emergency situation. Without loss of generality, we assume, for the sake of simplicity, that the above-explained negative role played by hospitals as contagion spreading factors, can be neglected here. Thus, define such a simplified notion of risk as $R: (0,1) \times \mathbb{R} \to (0,1)$ be a $C^2$ function, determined by $x \in (0,1)$ and $b \in \mathbb{R}$, i.e., $R = f(x,b)$. Consistently with previous analysis, we assume that $\frac{\partial R}{\partial x} > 0, \frac{\partial R}{\partial b} > 0$ and since, the level of risk is subject to saturation, $\frac{\partial^2 R}{\partial x^2} < 0, \frac{\partial^2 R}{\partial b^2} < 0$.

The level of risk is the target variable that the policy-maker tries to minimize, given the constraint constituted by the current carrying capacity, i.e., the endowment of hospital beds ($H$) financed by the expenditure in the healthcare system $G_H$. In principle, it may be considered as a dynamic variable, but we will proceed with a comparative-static analysis, by presenting policy intuitions from different initial configurations. Thus, the constraint is the result of the political orientation of the Government. In particular, let the part of the global allowance dedicated to intensive care beds, $HH$, be the sole remedy to the epidemy and define $HH = \alpha H, \alpha \in (0,1)$. The current carrying capacity of the healthcare system, i.e. the available number of hospital beds, is the function $Z: (0,1)^2 \times \mathbb{N} \times \mathbb{R} \to (0,1)$, defined as $Z = H - h \, x \, n$, where $h = (1 - \alpha)$ and $n$ is the population of the district under consideration. In per capita terms, it can be rewritten as $z = z_H - hx$, with $z = Z/n$ and $z_H = H/n$. Within a comparative statics perspective, for any given couple $(H, \alpha)$ we can consider a reduced form of the carrying capacity constraint, depending on the infection diffusion rate as a negatively sloped line in the plane (infection ratio, hospital beds), where also the convex contours of the risk function can be considered. It is worth to notice that the model refers to the variable hospital beds in both demand, $b$, and supply, $z$, terms. All in all, the proposed framework matches *required* and *available* beds per infection ratio. Policies may try to affect $x$ and can effectively tune $z$, by adjusting $\alpha$ and $G_H$, in such a way that the exploitation of available resources allows the minimization of $b$, at least, a tangency point between the capacity constraint and the risk level. The model assumes that the intensive care is the sole remedy to the epidemy. The fact that other effective protocols exists may have an effect on the slope of the linear constraint that represents the current carrying capacity of the healthcare system. In case other protocols exist, the model operates as described, but the line is translated downwards, other things being equal. In effect, other therapies or remedies would operate as an alternative to hospital beds, thus making the constraint less binding, i.e., reducing the needed allowance to face a given infection ratio.

# Acknowlegments

AEB, AP and  AR acknowledge financial support of the national project PRIN 2017WZFTZP *Stochastic forecasting in complex systems*. AP, AR, GR GI and VL acknowledge financial support of the fund "Linea Intervento 2 - Piaceri" from Catania University. The authors would like to thank Christian Mulder for his valuable comments and suggestions.

# Author's contribution statement

A.P., A.E.B, G.I., R.L., V.L., A.R., G.R. conceived and wrote the paper, collected and analyzed data. N.G. and C.Z. analyzed data and prepared figures. All authors reviewed the manuscript.